\documentclass{ws-procs9x6}

\newcommand{\beqn}{\begin{eqnarray}}
\newcommand{\eeqn}{\end{eqnarray}}

\newcommand{\dd}{{\mbox d}}
\newcommand{\eq}[1]{(\ref{#1})}

\begin{document}

\title{
\vskip -20mm
\rightline{\textmd{ITEP-LAT/2003-10}}
\rightline{\textmd{KANAZAWA-03-14}}
\vskip 5mm
%{\large January 10, 2003}
%\end{flushright}
%\vspace{1.0cm}
Gluodynamics in External Field: \newline A Test of the Dual Superconductor Picture
\footnote{\uppercase{T}alk given
at ``\uppercase{C}onfinement \uppercase{2003}'', \uppercase{RIKEN}, \uppercase{W}ako,
\uppercase{J}apan, 21-24 \uppercase{J}uly 2003.}
$^,$\footnote{\uppercase{T}his work is supported by the \uppercase{JSPS}
\uppercase{F}ellowship \uppercase{N}o. \uppercase{P}01023.}
}

\author{
M. N. Chernodub}
\address{
ITP, Kanazawa University, Kanazawa, 920-1192, Japan \\
ITEP, B. Cheremushkinskaya 25, Moscow, 117259, Russia}

\maketitle

\abstracts{We study gluodynamics in an external Abelian electromagnetic
field within the dual superconductor approach. We show that the
$SU(2)$ gluodynamics should possess a deconfining phase transition
at zero temperature at certain value of the external field. A dual
superconductor model for the $SU(3)$ gauge theory in external field
predicts a rich phase structure containing confinement, asymmetric
confinement and deconfinement phases. These results can be used to
check the validity of the dual superconductor description of gluodynamics
in external fields. We also discuss the gauge--independence of the obtained
results.}

{\bf 1.} We discuss the properties of the $SU(2)$ and $SU(3)$
gluodynamics in the external electromagnetic field using the dual
superconductor approach\cite{DualSuperconductor}
based on the Abelian monopole condensation. The condensate
-- observed\cite{MonopoleCondensation}  in various numerical simulations at low temperatures--
forces a chromoelectric flux coming from (anti-)quarks to squeeze into a confining string.
This picture of confinement has been confirmed in various lattice simulations\cite{Review}.

A common feature of known superconductors is that at strong enough magnetic fields
the superconductivity is destroyed and the superconductor goes in the normal (metal) state.
In the dual superconductor model a similar effect \cite{chernodub:external} leads to
the deconfinement phase transition. Below we study the phase diagram of gluodynamics
within the dual superconductor approach. We work in the Bogomol'ny limit\cite{Bogomolny:1975de}
supported by lattice simulations\cite{TypeI-II}.

{\bf 2.} Let us first consider the $SU(2)$ gauge
theory in the $4D$ Euclidean space. The infrared properties of the
vacuum of this model can be described by the Abelian Higgs
(or, Ginzburg--Landau) Lagrangian:
\beqn
L_{GL}[B,\Phi] = \frac{1}{4} F^2_{\mu\nu} +
\frac{1}{2} {\bigl| D_\mu(B) \, \Phi\bigr|}^2
+ \lambda {\biggl({|\Phi|}^2 - \eta^2 \biggr)}^2\,,
\label{Lagrangian:SU2}
\eeqn
where $F_{\mu\nu} = \partial_\mu B_\nu - \partial_\nu B_\mu$ is
the field strength of the dual gauge field $B_\mu$, $\Phi$ is the
monopole field with the magnetic charge $g_M$, and $D_\mu =
\partial_\mu + i g_M B_\mu $ is the covariant derivative. The
gauge field $B_\mu$ is dual to the third component of the gluon
field in an Abelian gauge. The model possesses the dual $U(1)$ gauge
symmetry, $B_\mu \to B_\mu - \partial_\mu \alpha$, $\Phi \to e^{i
g_M \alpha}\, \Phi$. The form of the potential implies the
existence of the monopole condensate, $|\langle\Phi\rangle| =
\eta>0$.

Consider the
four--dimensional sample of the (dual) superconductor occupying
half--space, $x_2 \geqslant 0$. Let us apply the constant external
EM field $F^{\mathrm{ext}}_{\mu\nu} = \varepsilon_{\mu\nu34}
H^{\mathrm{ext}}$ to the boundary of the superconductor.
The external field is screened due to the
induced superconducting current,
\beqn
J_\mu = \Im m \bigl(\Phi^* D_\mu(B) \Phi\bigr) \equiv |\Phi|^2
\cdot v_\mu\,, \quad  v_\mu = \partial_\mu \varphi + g_M B_\mu\,,
\label{current}
\eeqn
where we have set $\Phi = |\Phi| e^{i \varphi}$. The current is
parallel to the boundary of the superconductor. The monopole
kinetic term in Eq.~\eq{Lagrangian:SU2} can be written as $|D_\mu
\Phi|^2 = (\partial_\mu |\Phi|)^2 + |\Phi|^2 v^2_\mu$. Clearly, a
non--zero current provides an additional positively--defined term
in the Lagrangian ($\propto |\Phi|^2$). As a result, the external
field destroys the monopole condensate.

We treat the model \eq{Lagrangian:SU2} classically. It is convenient
to rewrite the action of the model~\eq{Lagrangian:SU2} as an
integral in a two-dimensional plane. One of the directions of the plane
is the depth of the dual superconductor, $x_2$,
while the second is given by the direction of the
current~\eq{current}. Choosing $J_\mu \propto \delta_{\mu,1}$
the first two terms of eq.~\eq{Lagrangian:SU2} become, respectively,
$F^2_{\mu\nu}/4  = H^2/2$ and
\beqn
{\bigl|D_\mu \Phi\bigr|}^2 = \sum\limits_{\alpha=1,2}
{\bigl|D_\alpha \Phi\bigr|}^2 =
{\biggl| \biggl(D_1 \pm i D_2 \biggr)\Phi\biggr|}^2
\mp 2 \varepsilon_{\alpha\beta} \partial_\alpha J_\beta \mp
g_M H \, {|\Phi|}^2\,.
\label{twostars}
\eeqn

In the Bogomol'ny limit, $g^2_M \slash \lambda = 8$, the action of the model
is:
\beqn
\!S \! = \!\frac{L_3 L_4}{2} \!\! \int \!\! \dd^2 x \Bigl\{
{\bigl| \bigl(D_1 - i D_2 \bigr)\Phi\bigr|}^2 +
{\Bigr[H + \frac{g_M}{2} \bigl(|\Phi|^2 - \eta^2
\bigr)\Bigr]}^2\Bigr\} + S_f + S_J,
\label{Lagrangian:SU2:3}
\eeqn
where $L_i$ is the (infinitely large) length of the dual
superconductor in $i^{\mathrm{th}}$ direction,
$S_J = L_3 L_4 \int \dd^2 x\, \varepsilon_{\alpha\beta} \partial_\alpha
J_\beta = L_1 L_3 L_4 J_1(x_2=0)$ is the action of the
surface current and $S_f = g_M L_3 L_4 \eta^2 \int \dd^2 x\, H \slash 2$
is the "flux" action.

The Bogomol'ny equations minimize the action~\eq{Lagrangian:SU2:3},
\beqn
(D_1 - D_2) \, \Phi = 0\,,\quad H + \frac{g_M}{2}\,
\Bigl(|\Phi|^2 - \eta^2\Bigr) = 0\,.
\nonumber
\eeqn
The second equation gives the monopole condensate at the boundary,
$|\Phi(x_2=0)|^2 = \eta^2 - 2 H^{ext} \slash g_M$.
The condensate disappears at the critical value of the external EM field,
$H^{\mathrm{ext}}= H_{\mathrm{cr}} = g_M \eta^2 \slash 2$.

At zero temperature and in the absence of the external fields
the tension of the string spanned on
trajectories of the fundamental charges can be evaluated
exactly in the Bogomol'ny limit\cite{Bogomolny:1975de,VeSc76}, $\sigma=\pi \eta^2$.
Using the Dirac relation between magnetic ($g_M$)
and electric ($g$) charges, $g_M g = 2 \pi$, we get the exact
value of the critical EM field in terms of the string tension:
\beqn
g H_{\mathrm{cr}} \slash \sigma = 1\,, \qquad \mbox{for SU(2)}\,.
\label{Hcr:SU2}
\eeqn
The dual superconductivity is destroyed in the {\it bulk}
if $H \geqslant H_{\mathrm{cr}}$.

\vskip 0.3cm {\bf 3.}
Now let us consider the Lagrangian of the ${[U(1)]}^2$ dual superconductor
model corresponding to $SU(3)$ gluodynamics\cite{MaSu81}:
\beqn
L = \frac{1}{4} F^a_{\mu\nu} F^{a,\mu\nu} +
\sum\limits^3_{i=1} \Bigl[\frac{1}{2}
{\bigl| (\partial_\mu + i g_M \varepsilon^a_i B^a_\mu) \Phi_i\bigr|}^2 + \lambda
{\Bigl({|\Phi_i|}^2 - \eta^2 \Bigr)}^2\Bigr]\,,
\label{Lagrangian:SU3}
\eeqn
where $F^a_{\mu\nu} = \partial_\mu B^a_\nu - \partial_\nu B^a_\mu$
is the field strength of the gauge field $B^a_\mu$, $a=3,8$.
The phases of the monopole fields, $\Phi_i$, $i=1,2,3$,
satisfy the relation $\sum^3_{i=1} \arg \Phi_i =0$.
The $\epsilon$--symbols are the root vectors of the $SU(3)$
group:  $\vec \epsilon_1=(1, 0)$, $\vec \epsilon_2=(-1 \slash 2,
-{\sqrt{3} \slash 2})$, $\vec \epsilon_3=(-{1 \slash 2}, {\sqrt{3}
\slash 2})$. In Eq.~\eq{Lagrangian:SU3} no summation over the Latin index $i$
is implied. The gauge fields $B^{3,8}_\mu$ are dual to the diagonal components
$a=3,8$ of the gluon field $A^a_\mu$. Lagrangian
\eq{Lagrangian:SU3} respects the dual $[U(1)]^2$ gauge invariance:
$B^a_\mu \to B^a_\mu + \partial_\mu \alpha^a$, $\theta_i \to
\theta_i + g_M (\varepsilon^3_i \alpha^3 + \varepsilon^8_i
\alpha^8)$, $a=3,8$, $i =1,2,3$, where $\alpha^3$ and $\alpha^8$
are the parameters of the gauge transformation.

In the ${[U(1)]}^2$ Bogomol'ny limit, $g^2_M \slash \lambda = 16 \slash 3$,
the equations of motion
\beqn
\Bigl(D^{(i)}_1 \pm i D^{(i)}_2 \Bigr)  \chi_i = 0\,, \quad
H^{(i)} \mp \frac{3 g_M}{4} \Bigl({|\chi_i|}^2 - \eta^2
\Bigr) = 0\,;\quad i =1,2,3\,,
\label{EOM}
\eeqn
give the same (due to the Weyl symmetry\cite{Weyl1,Weyl2})
critical value for all components of the EM field,
$H^{(i)} = \sum_{a=3,8} \epsilon^a_i H^a$, $i=1,2,3$:
\beqn
g H^{(i)}_{\mathrm{cr}} \equiv g {\tilde{H}}_{cr} \slash \sigma = 3 \slash 4\,,
\qquad \mbox{for SU(3)}\,.
\label{Hcr:SU3}
\eeqn
Here we used the Dirac condition and
the relation for the fundamental string tension, $\sigma = 2 \pi \eta^2$,
valid in the Bogomol'ny limit of the $[U(1)]^2$ model\cite{Chernodub:1999xi,Weyl1}.
\begin{figure*}[htb]
  \begin{center}
      \epsfxsize=10.0cm \epsffile{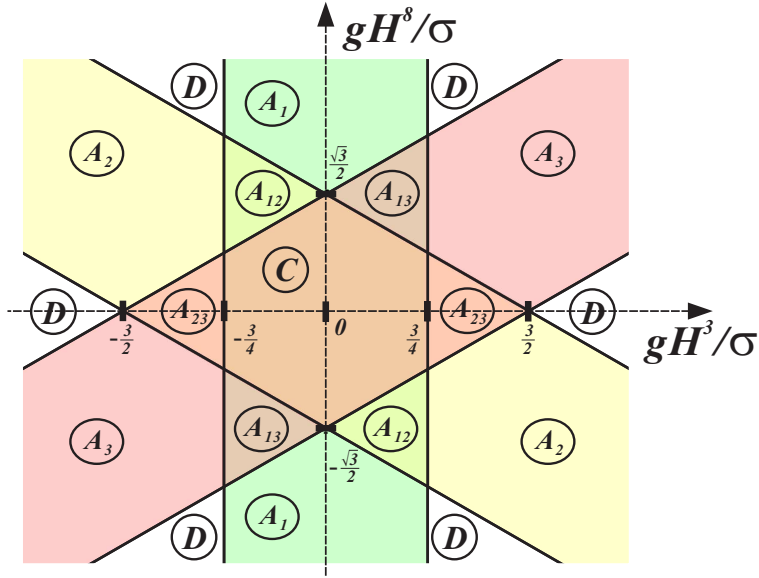}
  \end{center}
  \vspace{-0.5cm}
  \caption{The zero temperature
  phase diagram of the dual ${[U(1)]}^2$ Higgs model in the Bogomol'ny limit
  in the presence of the external electromagnetic field.}
  \label{fig:phase}
\end{figure*}
%\vskip -5mm
%
When the strength of the EM component $H^{(i)}$ reaches the
${\tilde H}_{cr}$ value then the monopole field, $\Phi_i$,
vanishes. Expressing the auxiliary fields $H^{(i)}$ in
terms of the components the EM field, $H^{3,8}$, and using
Eqs.(\ref{Hcr:SU3}) we get the phase diagram
of the ${[U(1)]}^2$ dual superconductor depicted in Figure~\ref{fig:phase}.

The phase diagram contains confinement ($C$), deconfinement ($D$)
and the asymmetric confinement phases ($A$). The phase
transition depends not only on the absolute value of the EM field
but also on the ("color") orientation of this field in the Cartan
subgroup. At low values of the field the model is always confining
regardless of the color orientation. However, as the absolute
value of the field is increased, the model enters -- depending on
color orientation -- one of six ($A_{12}$, $A_{13}$, $A_{23}$,
$A_{1}$, $A_{2}$ or $A_{3}$) asymmetric confinement phases. In the
$A_{ij}$ phase the $i$th and $j$th components of the monopole
field are condensed while the expectation value of the third
component is zero. In the phase $A_i$ the $i$th component is
condensed while the others two components are not. With
further increase of the field the model either enters the
deconfinement phase, $D$, or stays in one of the three
asymmetric confinement phases, $A_{1}$, $A_{2}$ or $A_{3}$.

One can show\cite{chernodub:external} that three of six
asymmetric confinement phases ($A_{12}$, $A_{13}$ and $A_{23}$) contain
one baryon and three meson states. These phases are
confining since the quarks of all three colors are
confined. The other three asymmetric confinement phases ($A_1$, $A_2$ and $A_3$)
contain only two light meson states while the baryon bound state is
absent.
%Quarks of a particular (phase--dependent) color are not confined in these phases.

\vskip 0.3cm {\bf 4.} Our results were obtained for the Abelian external
fields which are applied to the dual superconductor corresponding to a {\it fixed}
Abelian projection\footnote{Note that here we have implicitly
assumed that we are working in the Maximal Abelian projection where the
Bogomol'ny limit is realized\cite{TypeI-II}.}. Although the fact of the monopole condensation seems to be
projection--independent\cite{MonopoleCondensation}, important properties
of the dual superconductor may depend on the on chosen
Abelian projection\cite{Controversy}${}^,$\footnote{In particular,
this implies that results of Refs.\cite{CeaCosmaiRecent} for the phase diagram in the external Abelian fields can
not be compared with our predictions.
% because these results were obtained without the gauge fixing.
}.

We propose to investigate numerically the phase diagrams of the $SU(2)$ and $SU(3)$ gluodynamics
in the Maximal Abelian projection and compare them with the predictions of Ref.~\cite{chernodub:external}.
This comparison should reveal whether the dual superconductor picture works at strong external fields or not.

\end{document}